\documentclass[preprint]{aastex}

\shorttitle{SR 24 disks}
\shortauthors{Andrews \& Williams}

\begin{document}

\title{Submillimeter Array Observations of Disks in the SR 24 Multiple Star System\footnote{The Submillimeter Array is a joint project between the 
Smithsonian Astrophysical Observatory and the Academia Sinica Institute of 
Astronomy and Astrophysics, and is funded by the Smithsonian Institution and 
the Academia Sinica.}}

\author{Sean M. Andrews \& Jonathan P. Williams}

\affil{Institute for Astronomy, University of Hawaii, 2680 Woodlawn Drive, Honolulu, HI 96822}
\email{andrews@ifa.hawaii.edu, jpw@ifa.hawaii.edu}

\begin{abstract}
We present high-resolution aperture synthesis images from the Submillimeter 
Array of the 225\,GHz (1.3\,mm) continuum and $^{12}$CO $J=2-1$ line emission 
from the disks around components of the hierarchical triple system SR 24, 
located in the Ophiuchus star-forming region.  The most widely separated 
component, SR 24 S (with a projected semimajor axis $a = 832$\,AU), has a 
circumstellar disk with properties typical of those around single T Tauri 
stars.  The binary SR 24 N ($a = 32$\,AU) is undetected in the continuum, 
but has strong, resolved CO emission which likely originates in a circumbinary 
disk with a central gap.  The data constrain the total disk mass in the SR 24 N 
system to be $\lesssim 10^{-3}$\,M$_{\odot}$ and indicate that the depletion of 
CO onto dust grains is not more than 100 times larger than the mean value in 
the interstellar medium.  The SR 24 N disk is unusual in that it is only 
detected in line emission.  It is possible that other low mass disks around 
binaries and single stars may have been missed in single-dish continuum 
surveys.  
\end{abstract}

\section{Introduction}
Much progress has been made in constraining the basic physical properties of 
circumstellar disks with the inclusion of observations at (sub)millimeter 
wavelengths, beginning with a set of large 1.3\,mm continuum surveys 
\citep{bscg90,am94,osterloh95}.  At these long wavelengths, the thermal 
continuum emission from the dust disk is assumed to be optically-thin and thus 
provides a direct constraint on the total particle mass \citep{beckwith86}.  
Moreover, the (sub)millimeter spectrum is rich with rotational transitions of 
various molecules, presenting opportunities to study kinematics and chemistry 
\citep[e.g.,][]{qi03}.  Because an interferometer is required to disentangle 
the line emission from the disk and molecular cloud, only a handful of disks 
have constraints on the molecular gas phase.

The metamorphosis of a circumstellar disk into a planetary system can be 
inhibited if the local environment is hostile enough to severely disturb or 
destroy the disk.  A perhaps common example is photoevaporation in the vicinity 
of a massive star \citep[e.g.,][]{bally98}.  Another is dynamical disruption 
due to another star, either in a multiple system or a dense cluster 
environment.  The frequency of multiple systems among young stars is at least 
equal to that of main sequence stars in the field, and possibly significantly 
greater \citep{mathieu94,mathieu00}.  Numerical simulations demonstrate that 
the stability of circumstellar material in multiple star systems is jeopardized 
\citep{papaloizou77,lin93,artymowicz94}.  These studies show that, despite the 
dynamical interactions between disks and stars, individual circum\emph{stellar} 
disks can survive, albeit in truncated form, and large gaps would be produced 
in the circum\emph{binary} disks.  Single-dish continuum searches have 
confirmed that close binaries ($a \lesssim 50$\,AU) have significantly lower 
(sub)millimeter luminosities than wider binaries or single stars 
\citep{osterloh95,jensen94}.  However, the luminosities are still consistent 
with dynamically-carved gaps in circumbinary disks, and not necessarily the 
absence of disks altogether \citep{jensen96}.  Many young binaries have clear 
signatures of inner disks around at least one of the components 
\citep{jensen97}.  Recent evidence shows that widely-separated components are 
able to retain significant disks \citep{jensen03}.

The focus of this Letter is the triple system SR 24 (HBC 262, Haro 1$-$7), 
located in the Ophiuchus star-forming region \citep[$d = 
160$\,pc;][]{chini81}.  For convenience we refer to the primary as SR 24 S and 
the binary as SR 24 N.  The projected separation of SR 24 N/S is 5\farcs2 
at PA = 348\degr\ \citep{reipurth93b}.  SR 24 N itself is a binary with a 
projected separation of 0\farcs197 at PA = 87\degr\ \citep{simon95}.  Both SR 
24 N and S are known to be classical T Tauri stars with emission lines 
\citep{coku79}, Li absorption \citep{martin98}, and infrared excesses 
\citep{greene94}.  We present new high-resolution observations of the SR 24 
system in the 1.3\,mm continuum and CO $J=2-1$ line, and discuss their 
implications on understanding disk structure and planet formation in multiple 
star systems.  

\section{Observations and Data Reduction}
Millimeter interferometric observations of the SR 24 system were obtained on 
2004 August 2 and 21 with the Submillimeter Array \citep[SMA;][]{ho04} on Mauna 
Kea, Hawaii.  The data were taken with 7 antennas and double-sideband receivers 
with $\nu_{\rm{IF}}$ = 225.494\,GHz.  The $^{12}$CO $J=2-1$ line at 
230.538\,GHz was centered in the upper sideband.  The selected correlator setup 
gave 2\,GHz of continuum bandwidth in each sideband and a spectral resolution 
of 1.06\,km s$^{-1}$ ($\sim$0.8\,MHz) per channel.  During the observations, 
the zenith opacity was $\sim$0.12 and the median system temperature was 
$\sim$160\,K.  The combined compact (August 2) and extended (August 21) array 
configurations resulted in projected baselines from 9 to 186\,m.  Observations 
of SR 24 (with the phase center set at the position of SR 24 S: 
$\alpha$=16$^{\rm{h}}$26$^{\rm{m}}$58\fs5, 
$\delta$=$-$24\degr45\arcmin36\farcs67, J2000) were interleaved with two 
quasars (J1743$-$038 and NRAO 530) for use in gain calibration.  Additional 
observations of Jupiter and Uranus were obtained for bandpass and flux 
calibration.  The total on-source integration times for SR 24 were 1.7 and 3.3 
hours for the compact and extended configurations, respectively.  

The data were edited and calibrated using the IDL-based MIR software 
package.\footnote{http://cfa-www.harvard.edu/$\sim$cqi/mircook.html}  The 
bandpass response was determined from the Jupiter and Uranus observations.  
Complex gain calibration was conducted with both quasars, and the absolute flux 
calibration was set by J1743$-$038 ($F_{\nu} = 2.0$\,Jy).  The resulting quasar 
flux densities are in excellent agreement with several independent measurements 
conducted throughout 2004 August, from which we infer that the absolute flux 
calibration is accurate to $\sim$20\%.  The resulting rms phase noise for the 
quasars increased from $\sim$15\degr\ to 60\degr\ from the shortest to the 
longest projected baselines.  Standard imaging and deconvolution was conducted 
with the MIRIAD software package.  All maps were obtained using natural 
$uv$-weighting.  

\section{Results}

The millimeter continuum map of the SR 24 system has a FWHM resolution of 
$2\farcs3 \times 1\farcs2$ (beam PA = 43\degr) and rms noise level of 2\,mJy 
beam$^{-1}$.  A continuum source with a flux density of 68\,mJy is detected at 
the position of SR 24 S.  An elliptical gaussian fit to the source in a higher 
resolution map ($1\farcs4 \times 1\farcs0$, beam PA = 43\degr) made using only 
the extended array data shows that it is resolved, with a deconvolved FWHM size 
of $1\farcs23 \pm 0\farcs02 \times 0\farcs29 \pm 0\farcs01$ at PA = 28\degr.  
The SR 24 S flux density is a factor of 3 to 4 lower than previous single-dish 
estimates, where the FWHM beam sizes were $12-18$\arcsec\ 
\citep{reipurth93,am94,nurnberger98}.  The Fourier sampling of the SMA 
indicates that our sensitivity is decreased to only 10\% of the peak at those 
single-dish beam sizes \citep[e.g.,][]{wilner94}.  Therefore, the single-dish 
observations likely suffer contamination from extended emission due to 
distributed dust, e.g., an envelope around the entire SR 24 system.  No 
continuum emission is detected in the vicinity of SR 24 N to a 3$\sigma$ upper 
limit of 6\,mJy.  

Channel maps of the CO line at the same FWHM resolution as the continuum data 
have an rms noise level of 0.13\,Jy beam$^{-1}$ in each channel.  There is weak 
CO emission coincident with SR 24 S which has a spatially-averaged peak 
brightness temperature $T_b = 3$\,K at LSR velocity $v = 6$\,km s$^{-1}$.  
The velocity-integrated ($\Delta v = 2.1$\,km s$^{-1}$) line has an intensity 
$I_{\rm{CO}} \approx 4$\,K km s$^{-1}$.  A much stronger CO line is detected at 
the SR 24 N position, with a spatially-averaged peak $T_b = 5.7$\,K at $v = 
7$\,km s$^{-1}$ and a velocity-integrated intensity $I_{\rm{CO}} = 16.8$\,K km 
s$^{-1}$.  The line emission for SR 24 N is spatially resolved, and a fit with 
an elliptical gaussian gives a deconvolved FWHM size of $3\farcs2 \times 
1\farcs9$ at PA = 45\degr.  

Figure 1 shows the millimeter continuum map in grayscale, with overlaid 
contours of integrated CO emission.  The stellar positions have been determined 
from a $V$-band (F606W) image of SR 24 from the \emph{Hubble Space Telescope} 
(HST) archive.  While the positions of the stars relative to one another are 
known to very high accuracy, the absolute astrometric precision of the HST 
image is comparatively poor \citep[0\farcs7;][]{baggett02} due to the small 
field of view.  The astrometric accuracy of the SMA data was found to be 
$\sim0\farcs1$ by comparing the measured positions of the gain calibrators with 
their VLA coordinates.  The line and continuum peaks are thus at the positions 
of SR 24 N and S within the uncertainties.

\section{Discussion}
The disk masses in the SR 24 system can be computed by assuming that the 
millimeter continuum emission originates in an optically-thin, isothermal 
region with the interstellar gas to dust mass ratio ($\sim$100).  Using $T = 
30$\,K and an opacity of 0.02\,cm$^2$ g$^{-1}$ \citep{hildebrand83}, the SR 24 
S disk has $M_{disk} = 0.01$\,M$_{\odot}$, which is roughly the mass obtained 
by augmenting the abundances of the planets in the solar system to cosmic 
values \citep[the minimum mass solar nebula; e.g., ][]{weidenschilling77}.  The 
mass and projected diameter (190\,AU) of the SR 24 S disk are consistent with 
typical single T Tauri stars \citep[e.g.,][]{dutrey96}.  With the same 
parameters, we can constrain the SR 24 N disk(s) to have $M_{disk} \le 9.0 
\times 10^{-4}$\,M$_{\odot}$ (3$\sigma$), which is roughly the mass of 
Jupiter.

However, the strong CO emission at the position of SR 24 N indicates that there 
is disk material around this binary.  The velocity gradient across the 
emission is low ($\Delta v \le 2$\,km s$^{-1}$), and therefore it is not likely 
to originate in a molecular outflow.  There are no bright peaks similar to the 
emission coincident with SR 24 N in the primary beam ($2\farcm5$) of the line 
maps, making the probability of a chance alignment of cloud material 
negligible.  There are slight position-velocity offsets in the CO emission 
which are consistent with Keplerian rotation centered on the binary, but data 
with higher spectral resolution would be required to address this 
quantitatively.  In most disks, the low energy rotational lines of CO are 
expected to be optically-thick \citep{beckwith93}.  Therefore, we can derive a 
lower limit on the H$_2$ gas mass by treating the line as if it were 
optically-thin, following the method of \citet{scoville86}.\footnote{Here we 
also assume a 10\% He fraction and that the excitation temperature of the gas 
and the thermal temperature of the dust grains are equal.}  For $T=30$\,K, the 
SR 24 N disk has a gas mass $M_g \ge 2.8 \times 10^{-9} 
X_{\rm{CO}}^{-1}$\,M$_{\odot}$, where $X_{\rm{CO}}$ is the CO abundance 
relative to H$_2$ ($X_{\rm{CO}} \equiv [ \rm{CO} / \rm{H}_2 ]$).  The upper 
limit from the dust and the lower limit from the gas imply $2.8 \times 10^{-9} 
X_{\rm{CO}}^{-1} <$ $M_{disk}$ (M$_{\odot}$) $< 9.0 \times 10^{-4}$, which 
gives a 3$\sigma$ lower limit on the CO abundance, $X_{\rm{CO}} > 3.1 \times 
10^{-6} = 3.1 \times 10^{-2} X_{\rm{CO,ism}}$, where $X_{\rm{CO,ism}} = 
10^{-4}$.  For temperatures between $T = 20 - 100$\,K, the limit on 
$X_{\rm{CO}}$ varies from $0.02 - 0.26 X_{\rm{CO,ism}}$, showing that the CO 
depletion is no more than 100 in the SR 24 N disk.  In fact, these measurements 
are consistent with there being no depletion compared to the ISM in the SR 24 N 
disk.  This contrasts with the SR 24 S disk where the ratio of the CO line 
intensity to the continuum flux density is more than a factor of 100 lower and 
the CO depletion is $\sim 10^3$ ($T = 30$\,K), similar to other circumstellar 
disks.  

The broadband spectral energy distributions (SEDs) of SR 24 N and S, shown in 
Figure 2, are similar shortward of 20\,$\mu$m.  The infrared data, taken from 
the work of \citet{greene94}, indicate that both components are typical Class 
II sources.  The infrared excesses out to 20\,$\mu$m coupled with the strong 
H$\alpha$ emission lines are clear indications that both sources harbor 
actively accreting disks out to radii of at least 1\,AU.  The semimajor axis of 
the SR 24 N binary is $a = 32$\,AU.  Dynamical simulations of binaries 
with disks show that the individual circumstellar disks will be rapidly 
truncated at a fraction ($\sim0.2 - 0.4$, depending primarily on the orbital 
eccentricity) of the semimajor axis \citep{artymowicz94}.  In the case of SR 24 
N, the truncation radius is estimated to lie between 6 and 12\,AU.  Assuming 
radial temperature profiles of typical optically-thick Class II disks 
\citep[e.g.,][]{bscg90}, the truncated SR 24 N circumstellar disks should be 
able to produce thermal excesses out to $\sim$100\,$\mu$m, which is consistent 
with the SED.

The same dynamical process which truncates the circumstellar disks in the 
binary will open a large gap in the circumbinary disk to a radius a factor of 
$2-3\times$ the semimajor axis \citep{artymowicz94}, meaning $\sim$60$-$100\,AU 
for SR 24 N.  The observed CO emission, which extends to a radius of 
$\sim$250\,AU, likely originates in the remnant outer portion of the SR 24 N 
circumbinary disk.  Dynamical theory would predict that the SR 24 N disk 
structure consists of at least one circumstellar accretion disk with radius 
$\lesssim$12\,AU and a circumbinary ring which extends out from $\sim$80\,AU.  
The observations confirm the presence of an inner disk and a large gas disk 
extending to 250\,AU, but cannot constrain a gap size if one indeed exists.  
Since the average density in a disk is expected to decrease with radius, the CO 
depletion may be less in the outer regions.  If the CO emission around SR 24 N 
arises in an outer circumbinary ring, this may explain the very different 
relative strengths of the line and continuum emission in the SR 24 N and S 
disks.  Because they are uncontaminated by emission from an inner disk, 
circumbinary structures like that around SR 24 N may be a good way to directly 
probe physical conditions in the outermost regions of disks. 

Assuming the SR 24 N system has a reasonably eccentric orbit ($e \sim 0.4$), 
the upper limit on the continuum flux density is marginally consistent with the 
simple disk + gap models computed by \citet{jensen96}.  Therefore, a gap in the 
circumbinary disk may be all that is required to account for the non-detection 
in the continuum.  If the standard form of the mass surface density ($\Sigma 
\propto r^{-3/2}$) is unaffected by the clearing of the gap in the circumbinary 
disk, the fraction of the total mass removed is roughly a factor of 2 
(independent of the normalization of $\Sigma$).  This implies that the 
``undisturbed" disk structure around SR 24 N was not very massive; at most a 
few times Jupiter's mass, and much less massive than the SR 24 S disk.  
Alternatively, the SR 24 N disk may have evolved faster than the SR 24 S disk.

A few other young multiple star systems have been observed with millimeter 
interferometers.  At first glance, one of the most interesting is the UZ Tau 
quadruple system, observed by \citet{jensen96b}.  The widely-separated 
primaries SR 24 S and UZ Tau E\footnote{This star is actually a spectroscopic 
binary \citep{mathieu96}, and thus is another example of a circumbinary disk.} 
both have roughly minimum mass solar nebula 
disks detected in the continuum and CO.  However, the $a \sim 50$\,AU binary UZ 
Tau W has weak, unresolved ($\lesssim 70$\,AU) continuum emission and no CO 
line emission associated with it: essentially the complete opposite of SR 24 
N.  The small apparent size of the UZ Tau W continuum emission led 
\citet{jensen96b} to conclude that individual circumstellar disks contributed 
the continuum emission, whereas the gas around SR 24 N most likely arises in a 
circumbinary disk.  At the other extreme is GG Tau A, which has an essentially 
identical semimajor axis ($a \sim 36$\,AU) as SR 24 N and similar signatures of 
inner disk accretion, but a very massive ($M_d \approx 0.1$\,M$_{\odot}$) 
circumbinary ring well detected in both line and continuum emission 
\citep{koerner93,dutrey94}.  The disk around the UY Aur binary ($a \sim 
120$\,AU) is an intermediate case with extensive, bright CO emission and 
comparatively little continuum emission \citep{duvert98}.  Relatively massive 
circumbinary disks are also found around spectroscopic binaries \citep[e.g., GW 
Ori; ][]{mathieu95} and other triple systems \citep[e.g., T Tau; 
][]{weintraub89}.  

There is obviously a range of disk structures possible around multiple star 
systems.  With so few objects observed at high resolution, the cause of this 
diversity remains unclear.  The differences could be due to the evolutionary 
state of the system, or simply the (presumed) variety in the orbital parameters 
of the stellar components.  The strong CO detection of the SR 24 N disk in the 
absence of continuum emission introduces the intriguing possibility that other 
low mass disks, around both single and multiple stars, may have been missed in 
even the most sensitive single-dish continuum searches to date.  In spite of 
that possibility, the low disk masses inferred from (sub)millimeter 
observations around most binaries with separations similar to SR 24 N suggest 
that planet formation would not be possible given our current theoretical 
understanding.  On the contrary, the more widely-separated components in 
multiple star systems, like SR 24 S, appear to harbor disks which are just as 
capable of forming planetary systems as single stars 
\citep[see also][]{jensen96,jensen03}.

\section{Summary}
We have presented high-resolution aperture synthesis images from the SMA 
interferometer of the disks in the SR 24 triple system in Ophiuchus.  The 
comparatively isolated component in the system, SR 24 S, harbors a typical 
circumstellar disk.  On the other hand, the binary SR 24 N is only detected in 
CO line emission and not the continuum.  We suggest that SR 24 N is surrounded 
by at least one circumstellar disk as well as a circumbinary gas disk, 
presumably with a dynamically-carved gap.  Little is known about the molecular 
gas phase in disks around single stars, let alone high-order multiple systems.  
Because multiple star systems like SR 24 are coeval, the time evolution factor 
is removed, so that the variety in the observed disk structures is likely due 
to the effects of the local environment.

\acknowledgments
We are grateful for the assistance and expertise of the SMA staff, and in 
particular for the advice of Alison Peck, David Wilner, Shigehisa Takakuwa, and 
Chunhua Qi.  We also thank an anonymous referee and Mike Liu for useful 
suggestions which improved this paper.  This work was supported by NSF grant 
AST-0324328.

\begin{figure}
\plotone{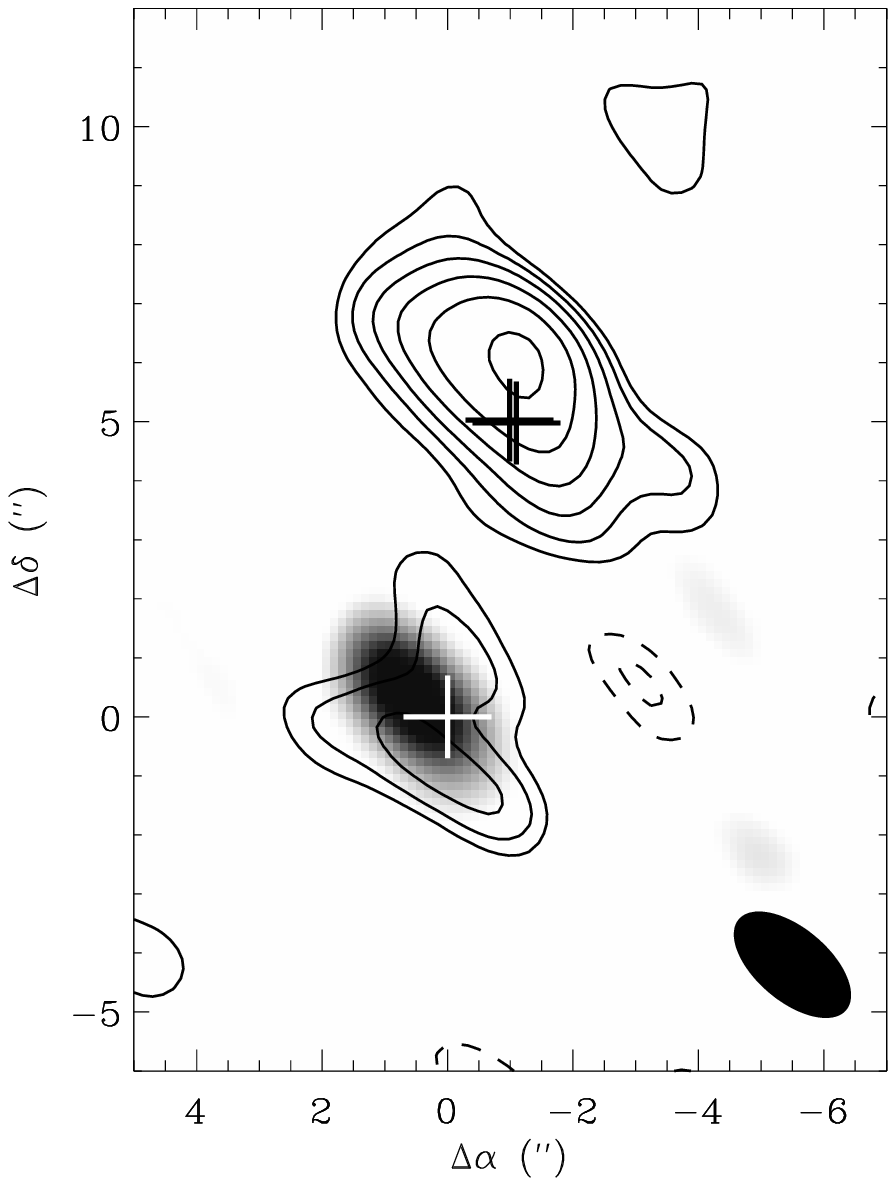}
\figcaption{Aperture synthesis images of the millimeter continuum (grayscale)
and velocity-integrated CO line emission (contours) in the SR 24 system.  The
CO contours begin at the 3$\sigma$ level (0.82\,Jy km s$^{-1}$) and each step
represents a factor of $\sqrt{2}$ in intensity.  Negative contours are dashed.  
The axes are offsets from the position of SR 24 S in arcseconds.  The FWHM 
synthesized beam size is shown in the lower right corner.  The stellar 
positions are marked with crosses of sizes indicating the uncertainty in the 
absolute HST astrometric positions and the SMA pointing.  Gas and dust are 
present in a circumstellar disk around SR 24 S, while only the gas phase of a 
circumbinary ring around SR 24 N is detected.}
\end{figure}

\begin{figure}
\plotone{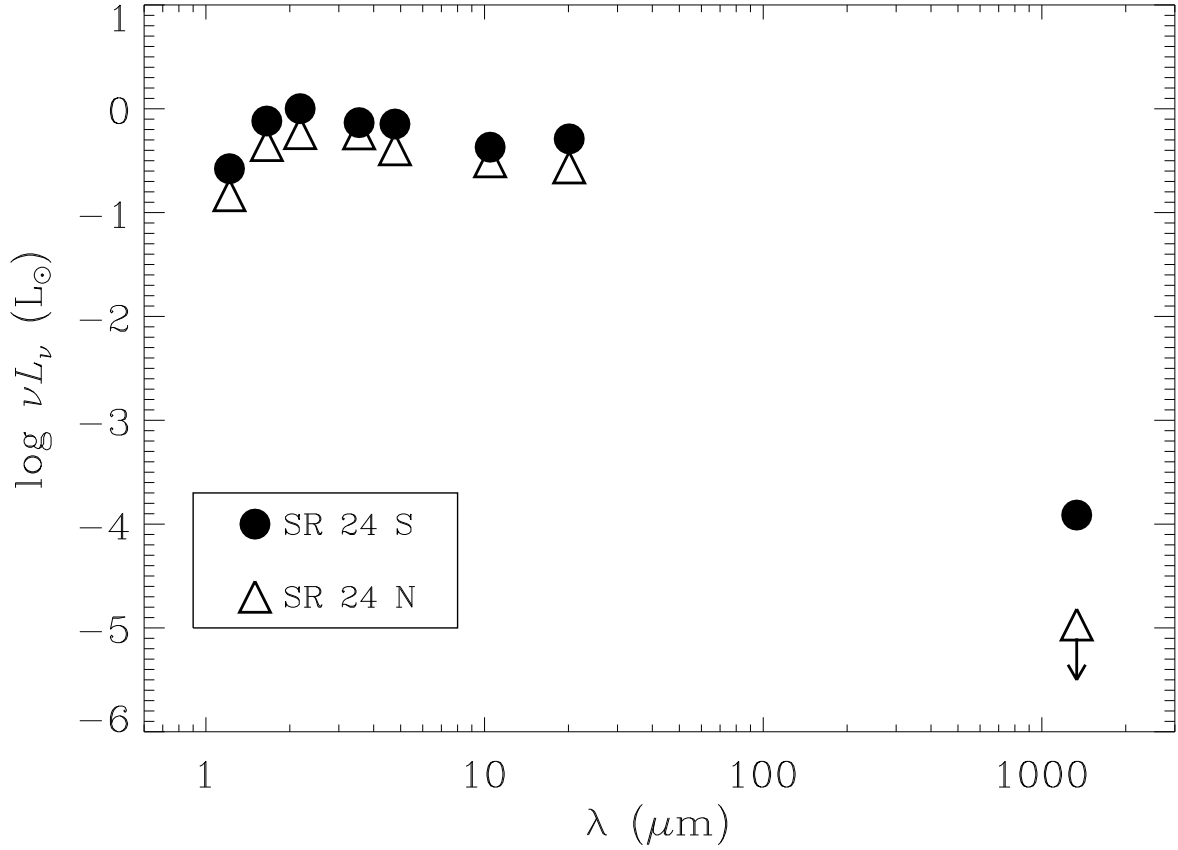}
\figcaption{The observed SEDs of SR 24 S (circles) and N (triangles).  The
infrared points are from \citet{greene94}.  The ordinate axis is defined as 
$\nu L_{\nu} = 4 \pi d^2 \nu F_{\nu}$.  Both SR 24 N and S have typical Class 
II infrared SEDs, suggesting the presence of inner disks.}
\end{figure}


\begin{thebibliography}{}
\bibitem[Andr\'{e} \& Montmerle(1994)]{am94} Andr\'{e}, P., \& Montmerle, T. 1994, \apj, 420, 837
\bibitem[Artymowicz \& Lubow(1994)]{artymowicz94} Artymowicz, P., \& Lubow, S. H. 1994, \apj, 421, 651
\bibitem[Baggett et al.(2002)]{baggett02} Baggett, S. et al. 2002, \emph{HST} WFPC2 Data Handbook, Version 4.0, ed. B. Mobasher (Baltimore: STScI)
\bibitem[Bally et al.(1998)]{bally98} Bally, J., Testi, L., Sargent, A., \& Carlstrom, J. 1998, \aj, 116, 854
\bibitem[Beckwith et al.(1986)]{beckwith86} Beckwith, S. V. W., Sargent, A. I., Scoville, N. Z., Masson, C. R., Zuckerman, B., \& Phillips, T. G. 1986, \apj, 309, 755
\bibitem[Beckwith et al.(1990)]{bscg90} Beckwith, S. V. W., Sargent, A. I., Chini, R. S., \& G\"{u}sten, R. 1990, \aj, 99, 924
\bibitem[Beckwith \& Sargent(1993)]{beckwith93} Beckwith, S. V. W., \& Sargent, A. I. 1993, \apj, 402, 280
\bibitem[Chini(1981)]{chini81} Chini, R. 1981, \aap, 99, 346
\bibitem[Cohen \& Kuhi(1979)]{coku79} Cohen, M., \& Kuhi, L. V. 1979, \apjs, 41, 743
\bibitem[Dutrey, Guilloteau, \& Simon(1994)]{dutrey94} Dutrey, A., Guilloteau, S., \& Simon, M. 1994, \aap, 286, 149
\bibitem[Dutrey et al.(1996)]{dutrey96} Dutrey, A., Guilloteau, S., Duvert, G., Prato, L., Simon, M., Schuster, K., \& M\'{e}nard, F. 1996, \aap, 309, 493
\bibitem[Duvert et al.(1998)]{duvert98} Duvert, G., Dutrey, A., Guilloteau, S., M\'{e}nard, F., Schuster, K., Prato, L., \& Simon, M. 1998, \aap, 332, 867
\bibitem[Greene et al.(1994)]{greene94} Greene, T. P., Wilking, B. A., Andr\'{e}, P., Young, E. T., \& Lada, C. J. 1994, \apj, 434, 614
\bibitem[Hildebrand(1983)]{hildebrand83} Hildebrand, R. H. 1983, QJRAS, 24, 267
\bibitem[Ho, Moran, \& Lo(2004)]{ho04} Ho, P. T. P., Moran, J. M., \& Lo, K.-Y. 2004, \apj, 616, L1
\bibitem[Jensen, Mathieu, \& Fuller(1994)]{jensen94} Jensen, E. L. N., Mathieu, R. D., \& Fuller, G. A. 1994, \apj, 429, L29
\bibitem[Jensen, Mathieu, \& Fuller(1996)]{jensen96} ------------ 1996, \apj, 458, 312
\bibitem[Jensen, Koerner, \& Mathieu(1996)]{jensen96b} Jensen, E. L. N., Koerner, D. W., \& Mathieu, R. D. 1996, \aj, 111, 2431
\bibitem[Jensen \& Mathieu(1997)]{jensen97} Jensen, E. L. N., \& Mathieu, R. D. 1997, \aj, 114, 301
\bibitem[Jensen \& Akeson(2003)]{jensen03} Jensen, E. L. N., \& Akeson, R. L. 2003, \apj, 584, 875
\bibitem[Koerner, Sargent, \& Beckwith(1993)]{koerner93} Koerner, D. W., Sargent, A. I., \& Beckwith, S. V. W. 1993, \apj, 408, L93
\bibitem[Lin \& Papaloizou(1993)]{lin93} Lin, D. N. C., \& Papaloizou, J. 1993, \emph{Protostars \& Planets III}, eds. E. Levy, \& M. S. Matthews (Tucson: Univ. Ariz. Press), 749
\bibitem[Mart\'{\i}n et al.(1998)]{martin98} Mart\'{\i}n, E. L., Montmerle, T., Gregorio-Hetem, J., \& Casanova, S. 1998, \mnras, 300, 733
\bibitem[Mathieu(1994)]{mathieu94} Mathieu, R. D. 1994, \araa, 32, 465
\bibitem[Mathieu et al.(1995)]{mathieu95} Mathieu, R. D., Adams, F. C., Fuller, G. A., Jensen, E. L. N., Koerner, D. W., \& Sargent, A. I. 1995, \aj, 109, 2655
\bibitem[Mathieu, Mart\'{\i}n, \& Magazz\`{u}(1996)]{mathieu96} Mathieu, R. D., Mart\'{\i}n, E. L., \& Magazz\`{u}, A. 1996, \baas, 128, 6005
\bibitem[Mathieu et al.(2000)]{mathieu00} Mathieu, R. D., Ghez, A. M., Jensen, E. L. N., \& Simon, M. 2000, \emph{Protostars and Planets IV}, eds. V. Mannings, A. P. Boss, \& S. S. Russell (Tucson: Univ. Ariz. Press), 703
\bibitem[N\"{u}rnberger et al.(1998)]{nurnberger98} N\"{u}rnberger, D., Brandner, W., Yorke, H. W., \& Zinnecker, H. 1998, \aap, 330, 549
\bibitem[Osterloh \& Beckwith(1995)]{osterloh95} Osterloh, M., \& Beckwith, S. V. W. 1995, \apj, 439, 288
\bibitem[Papaloizou \& Pringle(1977)]{papaloizou77} Papaloizou, J. C. B., \& Pringle, J. E. 1977, \mnras, 181, 441
\bibitem[Reipurth et al.(1993)]{reipurth93} Reipurth, B., Chini, R., Kr\"{u}gel, E., Kreysa, E., \& Sievers, A. 1993, \aap, 273, 221
\bibitem[Reipurth \& Zinnecker(1993)]{reipurth93b} Reipurth, B., \& Zinnecker, H. 1993, \aap, 278, 81
\bibitem[Scoville et al.(1986)]{scoville86} Scoville, N. Z., Sargent, A. I., Sanders, D. B., Claussen, M. J., Masson, C. R., Lo, K. Y., \& Phillips, T. G. 1986, \apj, 303, 416
\bibitem[Simon et al.(1995)]{simon95} Simon, M., et al. 1995, \apj, 443, 625
\bibitem[Qi et al.(2003)]{qi03} Qi, C., Kessler, J. E., Koerner, D. W., Sargent, A. I., \& Blake, G. A. 2003, \apj, 597, 986
\bibitem[Weidenschilling(1977)]{weidenschilling77} Weidenschilling, S. J. 1977, Ap\&SS, 51, 153
\bibitem[Weintraub, Masson, \& Zuckerman(1989)]{weintraub89} Weintraub, D. A., Masson, C. R., \& Zuckerman, B. 1989, \apj, 344, 915
\bibitem[Wilner \& Welch(1994)]{wilner94} Wilner, D. J., \& Welch, W. J. 1994, \apj, 427, 898
\end{thebibliography}
\end{document}